\documentclass[12pt]{article}

\usepackage{scicite}
\usepackage{times}
\usepackage{graphicx}
\usepackage{dcolumn}
\usepackage{bm}
\usepackage{amsmath,amssymb,amsfonts}
\usepackage{color}

\newenvironment{sciabstract}{%
\begin{quote} \bf}
{\end{quote}}

\newcounter{lastnote}
\newenvironment{scilastnote}{%
\setcounter{lastnote}{\value{enumiv}}%
\addtocounter{lastnote}{+1}%
\begin{list}%
{\arabic{lastnote}.}
{\setlength{\leftmargin}{.22in}}
{\setlength{\labelsep}{.5em}}}
{\end{list}}

\definecolor{orange}{rgb}{1,0.5,0}

\title{Three-Dimensional Anderson Localization of Ultracold Matter}

\author{S. S. Kondov,$^{1}$ W. R. McGehee,$^{1}$ J. J. Zirbel,$^{1}$ B. DeMarco$^{1\ast}$ \\
\\
\normalsize{$^{1}$Department of Physics, University of Illinois at Urbana-Champaign, Urbana, Illinois 61801, USA}\\
\\
\normalsize{$^\ast$To whom correspondence should be addressed; E-mail:  bdemarco@illinois.edu.}
}

\date{}

\begin{document}
\baselineskip24pt
\maketitle

\begin{sciabstract}

Anderson localization (AL) is a ubiquitous interference phenomenon in which waves fail to propagate in a disordered medium.  We observe three-dimensional AL of non-interacting ultracold matter by allowing a spin-polarized atomic Fermi gas to expand into a disordered potential.  A two-component density distribution emerges consisting of an expanding mobile component and a non-diffusing localized component.  We extract a mobility edge that increases with the disorder strength, whereas the thermally averaged localization length is shown to decrease with disorder strength and increase with particle energy. These measurements provide a benchmark for more sophisticated theories of AL.

\end{sciabstract}

\maketitle

Wave propagation in disordered media is affected by scattering from random impurities.  When those scattered waves self-interfere destructively, a phenomenon known as Anderson localization (AL) can arise \cite{Anderson1958}.  AL occurs in a wide variety of classical and quantum materials, impacting the transport of light \cite{Wiersma1997,PhysRevLett.96.063904}, acoustic \cite{Hu2008}, and matter \cite{Billy2008,Roati2008} waves.  Localization is known to affect electrical conductivity in solids as a result of scattering from impurities and defects \cite{Lee1985}, which is relevant to technological applications. We investigate 3D AL of a non-interacting ultracold atomic Fermi gas in a disordered potential created using optical speckle.  The behavior of the gas is qualitatively consistent with several features of 3D AL and is shown to be incompatible with simple trapping and classical diffusion.  Whereas in sufficiently large 1D and 2D systems particles can be localized no matter how weak the disorder, AL is not inevitable in 3D \cite{PhysRevLett.42.673}.  This distinctive feature of 3D AL---that the disorder strength sets a critical energy, the mobility edge, below which states are localized---is reflected here in measurements of the fraction of localized particles. We perform a direct measurement of how the mobility edge depends on disorder strength by smoothly tuning the speckle intensity.

We create trapped, ultracold gases of fermionic $^{40}$K atoms using standard techniques \cite{DFG1999,Esslinger1998,PhysRevA.79.063605}. The atoms are confined in an optical dipole trap, cooled to 170--1500~nK, and then spin polarized \cite{SOM}.  Quantum statistics do not play a significant role in the measurements discussed here because the lowest temperature we sample corresponds to roughly one-half of the Fermi temperature of the gas.  A focused optical speckle field (Fig. 1A) created by 532~nm light scattered through a diffuser and consisting of randomly distributed light and dark regions is superimposed on the trapped gas (as in \cite{White2009}).  The atoms experience a repulsive potential proportional to the speckle intensity, resulting in a disordered potential characterized by an approximately Gaussian autocorrelation function with $\zeta_x=270$~nm and $\zeta_z=1600$~nm r.m.s. radii. The laser beam creating the speckle propagates in the vertical $z$ direction; we refer to the transverse directions (residing in the focal plane of the speckle) as $x$.  The speckle intensity varies somewhat across the gas since it has a Gaussian envelope with a 170~$\mu$m $1/e^2$ radius along $x$ and $y$ and a 400~$\mu$m Rayleigh range along $z$.  The disorder strength $\Delta$, which is the potential energy averaged over $\zeta_x$ and $\zeta_z$ at the center of the speckle field, can be continuously varied from 0--1000~$k_B\times$nK by adjusting the 532~nm laser power ($k_B$ is Boltzmann's constant).  After the speckle field is slowly turned on over 200~ms, the trap is suddenly turned off, and the gas is allowed to expand in the disordered potential while supported against gravity by a magnetic field gradient.  Until the expansion, the Gaussian momentum distribution of the trapped gas is unchanged by the presence of the speckle potential \cite{SOM}.

Our experiments are distinguished primarily in two ways from previous work on AL of ultracold atoms.  First, we work in 3D, where AL relies on small angle scattering rather than partial back reflections as in 1D. Also, we eliminate inter-particle interactions by using a spin-polarized gas of fermionic atoms at temperatures far below the $\sim150$~$\mu$K $p$-wave collision threshold \cite{DeMarco1999}.  In previous experiments, bosonic atoms were employed and the effects of interactions were suppressed or eliminated by using a Feshbach resonance \cite{Roati2008} or by reducing the density \cite{Billy2008}.

AL in 3D is conditional on the Ioffe-Regel criterion, which is equivalent to the quantum wavelength of the particle exceeding the Boltzmann transport mean-free path $\ell_B$.  For the maximum $\Delta$ we achieve and the range of particle energies we sample, $\ell_B$ reaches a lower limit set by the speckle correlation length \cite{Kuhn2007}.  Using $\ell_B\approx\left(\zeta_x^2\zeta_z\right)^{1/3}$ and the thermal deBroglie wavelength $\Lambda_{dB}=h/\sqrt{2\pi m k_B T}$, the Ioffe-Regel criterion corresponds to $T\lesssim300$~nK.  Here, $h$ is Planck's constant, $T$ is temperature, and $m$ is the atomic mass.  Since a spread of particle wavelengths are present in the gas and the Ioffe-Regel criterion is not a precise constraint, localization is possible even for temperatures somewhat above this limit.

We probe localization by imaging the density profile after the gas has expanded for a variable time in the speckle potential.  As observed for the typical absorption image (Fig. 1B), a two-component profile emerges for any finite disorder energy.  The mobile component has a Gaussian profile similar to that of a freely expanding gas, but expands (at a constant velocity) more rapidly than a thermal gas.  In stark contrast, the stationary localized component has a profile along $z$ that is approximately exponential, and a distribution along $x$ that is well described by a Gaussian.  Although strongly localized single particles are known to generally possess exponential density profiles \cite{Lee1985} with localization lengths $\xi$ that depend on the particle energy $E$, $\Delta$, and the microscopic disorder parameters \cite{Kuhn2007}, a theoretical distribution applicable to our experiment (e.g., accounting for a thermal average over particle energies and localization lengths) is unresolved.  Density profiles are therefore analyzed using a heuristic fit that reproduces their basic features \cite{SOM}. The dynamics of the localized component are measured (after subtracting the mobile component from images \cite{SOM}) using a fit to a distribution proportional to $e^{-x^2/2\sigma_x^2-|z|/\xi_z}$ with an exponential localization length $\xi_z$ and r.m.s. radius $\sigma_x$ along $x$ (Fig. 1C,D).

As shown in Fig. 2, the size of the localized component becomes fixed after it rapidly expands along $z$ for $\sim25$~ms; the transverse size is approximately constant at the in-trap size. This apparent lack of diffusion cannot be explained classically.  In a 3D speckle field there are no local intensity minima that can trap and classically localize particles, in contrast to 1D \cite{PhysRevLett.95.070401,PhysRevLett.95.170409} and 2D \cite{PhysRevLett.104.220602,Pezze2011}. Rather, a 3D speckle field consists of a rich collection of topological features, such as dark optical vortex rings and lines that do not propagate in a single direction but instead wander in all three directions on the length scales associated with $\zeta_x$ and $\zeta_z$ \cite{O'Holleran2008}. While the vortex rings can trap particles in a finite volume, calculations of the percolation threshold for a 3D speckle field establish that less than 0.2\% of the particles are classically trapped for all of the data presented here \cite{Pilati2010}. The expansion of the localized component is also inconsistent with classical dynamics. We numerically simulated classical trajectories in a 3D speckle potential for a thermal ensemble of particles under the conditions used for the data in Fig. 2C \cite{SOM}. The simulated sizes after expansion (solid lines) are incompatible with the observed dynamics of the localized component.

Based on the expansion dynamics, we interpret the localized component as being comprised of particles with energies below the mobility edge $E_c$, and atoms with higher energy constituting the mobile component.  The density profile of the localized component after expansion can be straightforwardly understood in this context.  The Gaussian distribution along $x$ is stationary because the transverse localization lengths are much smaller than the initial size of the gas; the observed profile cannot be explained by diffusion strongly suppressed via localization solely in $z$ \cite{SOM}.  The profile along $z$ results from a thermal average of exponentially localized wavefunctions with much longer localization lengths, distributed such that the overall profile is approximately exponential.  A disparity in localization lengths between $x$ and $z$ is expected because $\ell_B$, which controls the localization length and the range of energies over which it diverges when $E\sim E_c$, strongly depends on the disordered potential correlation length.  At low energies, $\ell_B\propto\zeta^2$, and at high energies $\ell_B\propto\zeta^5$ (in the weak scattering limit) \cite{Kuhn2007}, and therefore the typical localization length should be at least 36 times larger along $z$.

By measuring the fraction of localized particles, we determine the mobility edge that separates localized from extended states. The mobility edge we observe is unrelated to a similar quantity in 1D \cite{Billy2008} that results from correlations in the disordered potential \cite{PhysRevA.80.023605}. In 1D, an effective mobility edge arises because the quantum scattering generating back reflections becomes higher order above a momentum cutoff determined by the Fourier spectrum of the speckle potential.  We measure the localized fraction after the atoms are held in the speckle potential for 20~ms to resolve the mobile component and to minimize the impact of decay of atoms from the localized component (evident in Fig. 2A,B). This decay to finite localized fraction at long times is characterized by a 30--50~ms exponential time constant, does not strongly depend on temperature, and may in part arise from the atoms sampling lower intensity regions away from the center of the speckle field.  Figure 3A shows the fraction of localized atoms determined from fits to density profiles \cite{SOM} for temperatures spanning 200--1500~nK and across the full range of accessible disorder energy.  More particles are localized as $\Delta$ is increased and $E_c$ correspondingly grows, or as $T$ is decreased and fewer particles are thermally excited above $E_c$.

The mobility edge is determined from each point in Fig. 3A by calculating the momentum cutoff $\sqrt{2mE_c}$ required to achieve the measured fraction of localized particles for a 3D spherically symmetric Gaussian momentum distribution consistent with the temperature \cite{SOM}.  Figure 3B shows $E_c$ averaged across data taken at different temperatures for a fixed $\Delta$.  While $E_c$ increases with $\Delta$, it does not follow the prediction that $E_c\propto\Delta^2$ from the self-consistent Born approximation \cite{Yedjour2010} and weak-scattering theory \cite{Kuhn2007}; for $\Delta\geq80\:k_B\times$nK the data fit well to a power law with $E_c\propto\Delta^{0.59\pm0.02}$ (dashed line in Fig. 3B).  Although these approaches are broadly applied to localization, this disagreement is likely a result of their failure precisely in the regime of AL.  How $\Delta$ affects $E_c$ has not been predicted for the strongly localized regime that we probe here.

The dependence of the measured localization length on the disorder strength and particle energy (controlled by changing temperature) is shown in Fig. 4.  While data are only shown for a limited range of $\Delta$ and $T$, the observed qualitative behavior---that $\xi_z$ increases with $T$ and decreases with $\Delta$---is characteristic of the entire range of parameters explored here. Extracting universal parameters from the data shown in Fig. 4 is difficult in the absence of theory because the measured quantities are averaged across all particle energies present in the gas.  The general monotonic trends, however, are consistent with a weak-scattering picture, in which the localization length is controlled by $\ell_B\propto\sqrt{E}/\Delta^2$ \cite{Kuhn2007} at low energy.

The variation of the speckle intensity across the gas is a complication that may affect the interpretation of the data shown in Figs. 3 and 4.  While we have used the central speckle intensity $\Delta$ as a single parameter to characterize the disorder strength in Figs. 3 and 4, the speckle envelope causes a range of speckle intensities to be sampled by the atoms.  The effect of the transverse speckle envelope is minor, since $\sigma_x\lesssim35$~$\mu$m, which is small compared with the Gaussian waist.  However, the maximum localization length in $z$ is approximately 270~$\mu$m, which leads to the disorder energy averaged across the atomic density distribution reduced to approximately 70\% of $\Delta$; for smaller $xi_z$ the disorder energy averaged in this way is closer to $\Delta$.  Although we have determined that this averaging does not directly affect the measured scaling of the mobility edge with disorder strength \cite{SOM}, the effect of the speckle envelope will likely be important to future comparisons between theory and our data.

In the future, the exquisite control possible over ultra-cold disordered gases may enable measurements to shed new light on aspects of 3D localization that are not well understood or are complicated by inter-particle interactions or dissipation in other systems (see \cite{SPreview} for a review).  An important issue that may be addressed if the single-particle states can be resolved (using Bragg spectroscopy, for example \cite{PhysRevLett.101.135301}) is the critical exponent that controls how the localization length diverges for energies near the mobility edge.  The influence of inter-particle interactions on the localization of fermions---another crucial question---can be studied by controllably introducing a second spin species; of particular interest is the impact of disorder on the BEC-BCS crossover (see \cite{PhysRevLett.99.250402,Dey2008} and references therein).  Finally, the impact of finite correlations in the disordered potential may be investigated using simple Gaussian or more complex holographic optics \cite{Pasienski:08}.
\\

\clearpage
\pagebreak
\begin{figure}
\centering
\includegraphics[width=10cm]{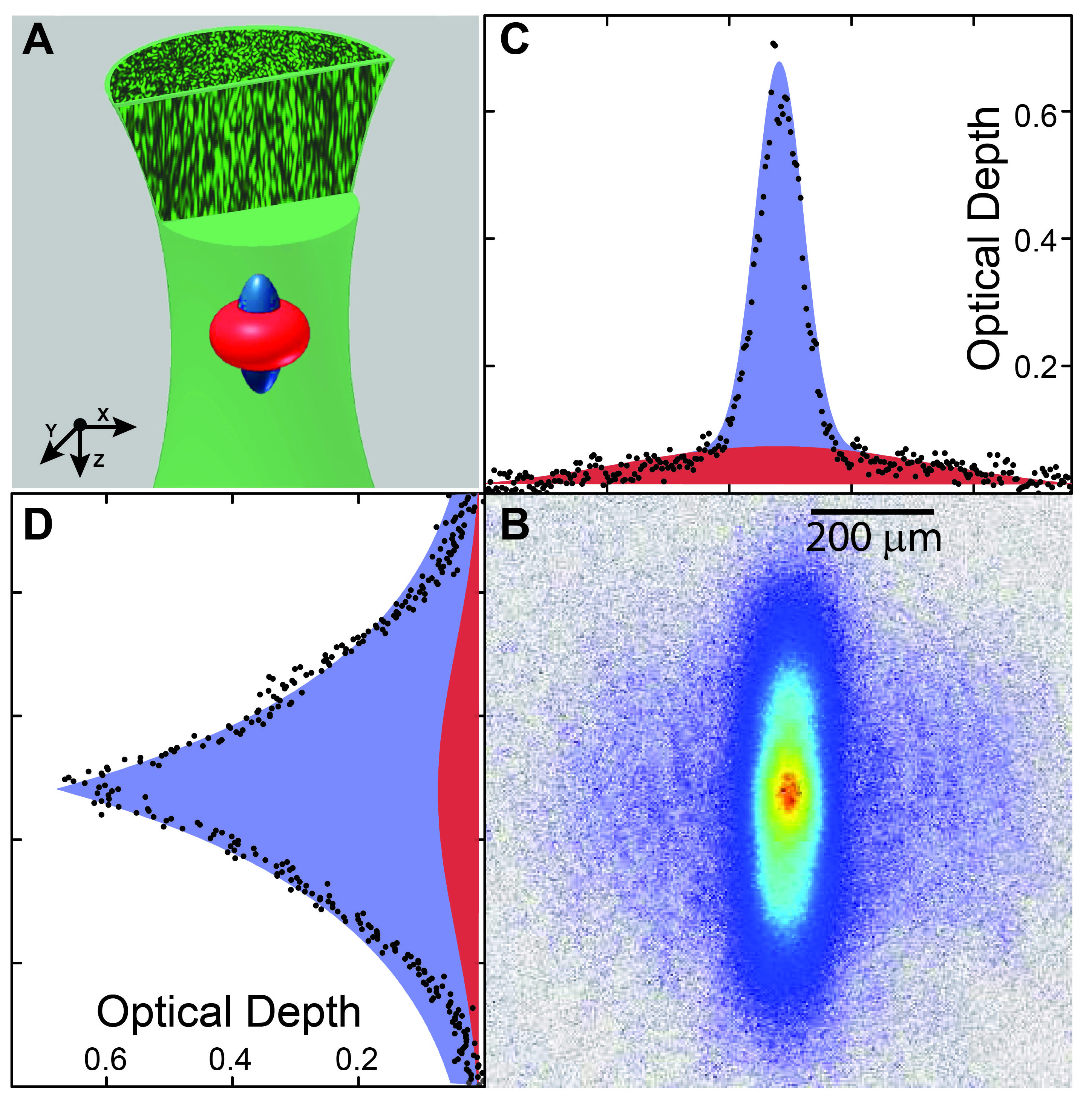}
\caption{Ultracold gas expanding into an optical speckle field (green) and separating into localized (blue) and mobile (red) components.  ({\bf B}) The measured optical depth, proportional to the atomic density integrated through $y$, is shown in false color.  The image is taken of a 480~nK gas that has expanded for 20~ms through the disordered potential with $\Delta=240\:k_B\times$nK. All images shown in this manuscript are averaged over at least 5 experimental realizations. Slices are shown through the image along $x$ ({\bf C}) and $z$ ({\bf D}). The filled curves are fits to independent mobile (red) and localized (blue) components.}
\end{figure}

\pagebreak
\begin{figure}
\centering
\includegraphics[width=10cm]{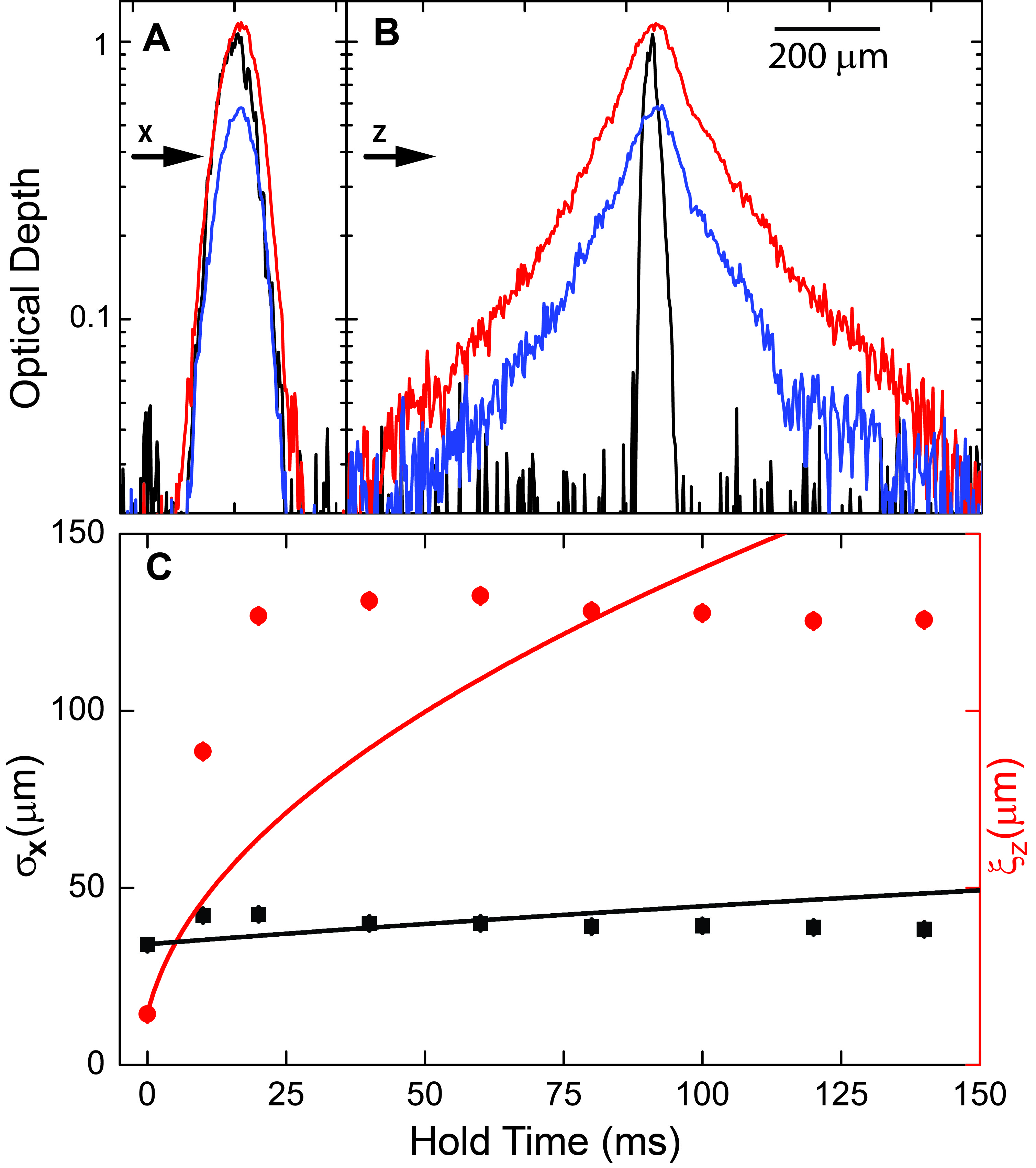}
\caption{Dynamics of the localized component for a 390~nK gas with $\Delta=600\:k_B\times$nK. Slices through an image taken before release from the trap (black) and after the gas has expanded for 40~ms (red) and 140~ms (blue) along the $x$ ({\bf A}) and $z$ ({\bf B}) directions. The wavelength of the imaging laser was changed to reduce the optical depth of the in-trap image by a factor of 15. The decrease in optical depth between 40~ms and 140~ms is a consequence of atoms slowly decaying from the localized component. ({\bf C}) The measured localization length $\xi_z$ (\textcolor{red}{$ \bullet $}) and r.m.s. size $\sigma_x$ ($\blacksquare $) of the localized component for variable hold time in the speckle potential.  Each point is determined from an average of 6 images; the error bars (not visible for every point) in all figures are the standard error unless otherwise specified.  The simulated classical expansion (solid lines) ignores rapid ballistic motion at short times, which lasts for several ms along $z$.}
\end{figure}

\pagebreak
\begin{figure}
\centering
\includegraphics[width=10cm]{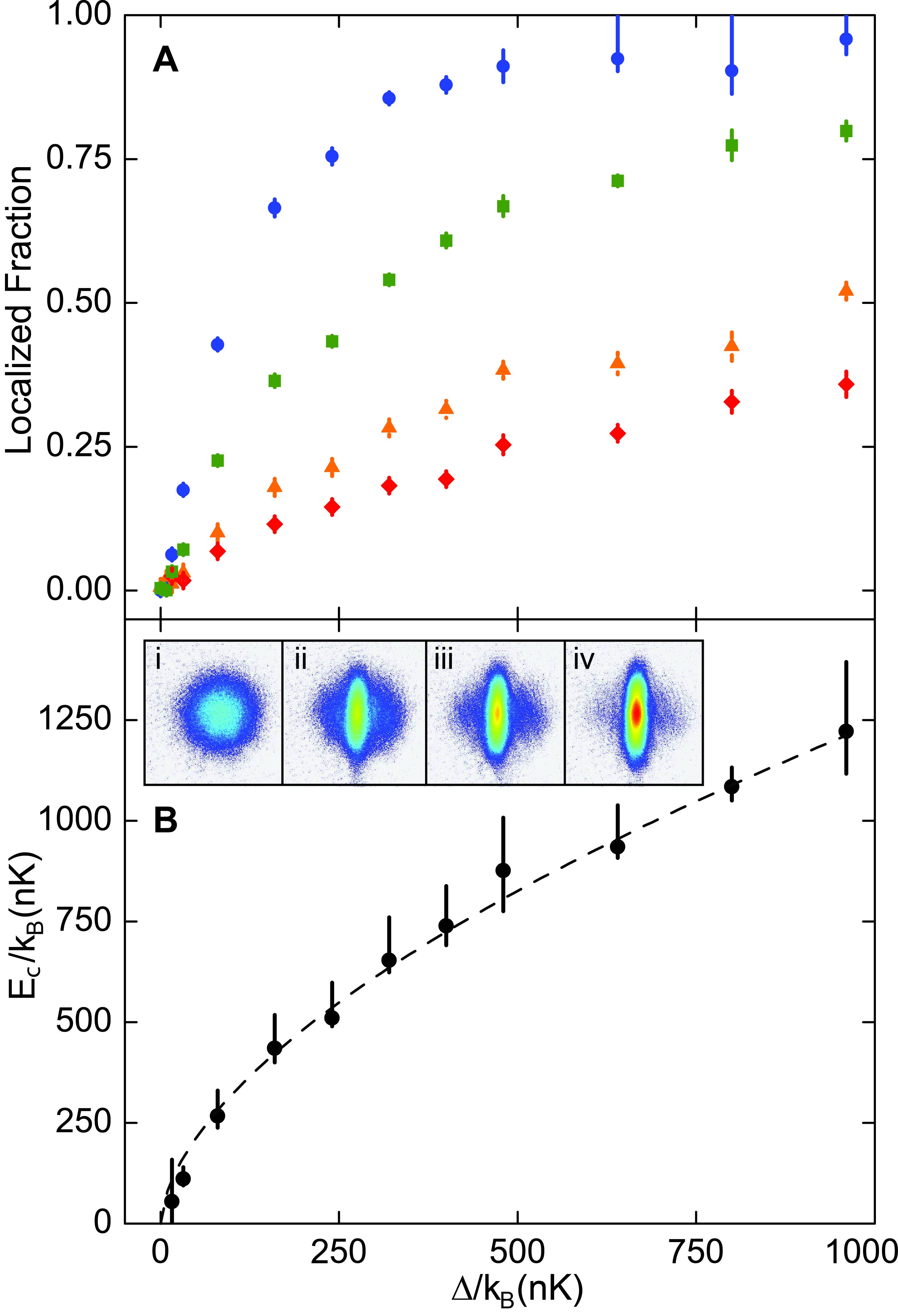}
\caption{ ({\bf A})  The fraction of atoms in the localized component measured after 20~ms of expansion into the disordered potential for varying $\Delta$ and $T=240 \pm 20$~nK (\textcolor{blue}{$ \bullet $}),  $480 \pm 20$~nK (\textcolor{green}{$ \blacksquare $}),  $1130 \pm 60$~nK (\textcolor{orange}{$ \blacktriangle $}), and  $1470 \pm 230$~nK (\textcolor{red}{$ \blacklozenge $}).  Each point is determined from fits to 5 images.  The growing localized fraction with increasing $\Delta$ is evident in the insets to ({\bf B}), which are images (with a false color logarithmic scale) taken at $T=480$~nK and $\Delta=0$ (i), 80 (ii), 160 (iii), and $320\: k_B\times$nK (iv).  ({\bf B}) Using the data in ({\bf A}), the mobility edge $E_c$ is determined at each $\Delta$.  Each point is a weighted average of $E_c$ accounting for the uncertainty in $T$ and localized fraction.  The error bars are the range of $E_c$ for the different temperatures that contribute to each point.}
\end{figure}

\pagebreak
\begin{figure}
\centering
\includegraphics[width=10cm]{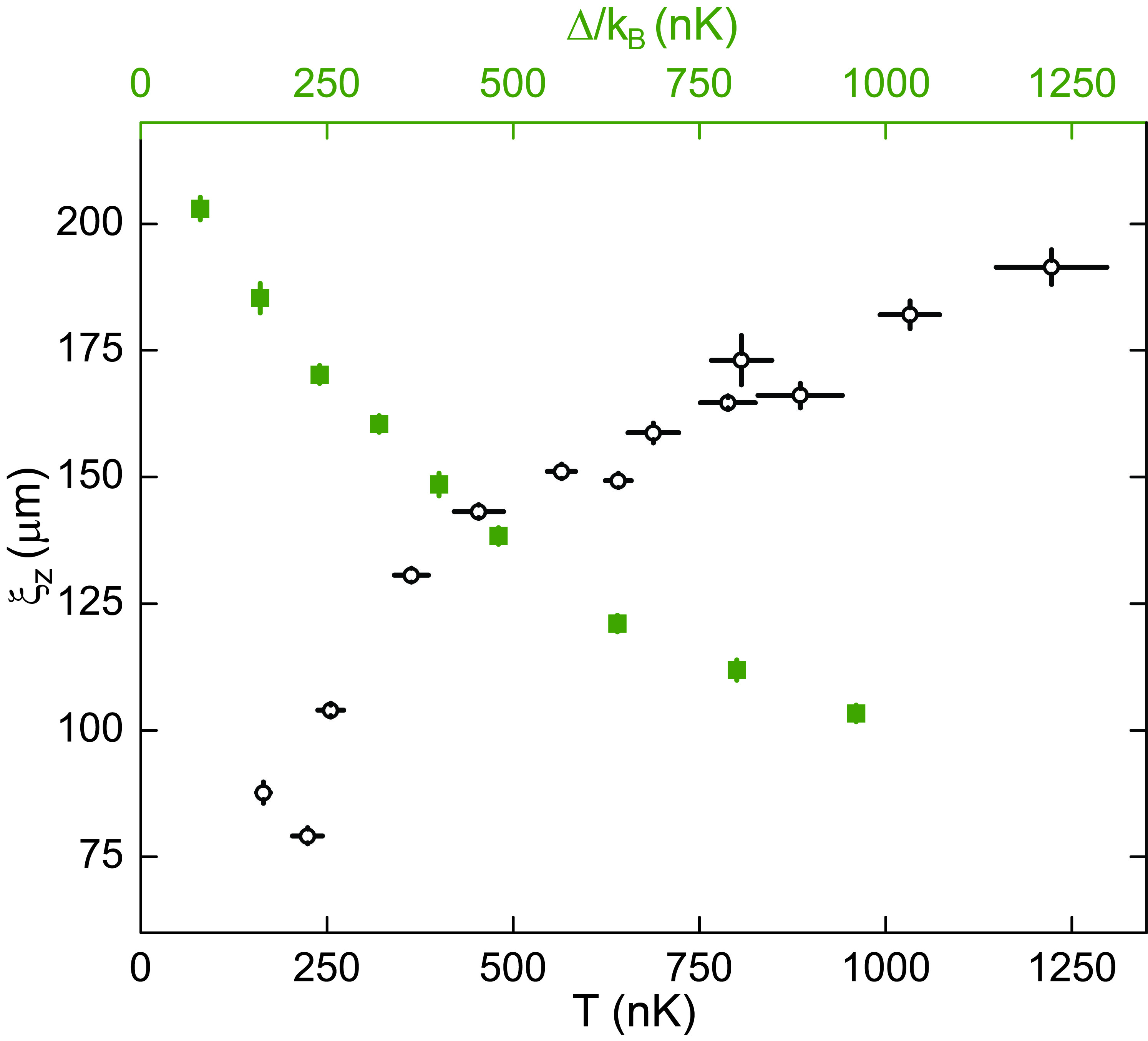}
\caption{The dependence of the measured localization length $\xi_z$ on disorder strength $\Delta$ at fixed temperature $T=480$~nK (\textcolor{green}{$ \blacksquare $}) and on $T$ at fixed $\Delta=480\:k_B\times$nK ($\bigcirc$).  The green points are from the same data set as in Fig. 3A, and the black points are an average of 10 experimental realizations.  The error bars in $T$ are from the uncertainty in the thermal expansion velocity.}
\end{figure}

\clearpage
\pagebreak
\section*{Supporting Online Material}
\subsection*{Materials and Methods}

We prepare ultracold gases of $^{40}$K atoms using techniques similar to those in \cite{DFG1999} and \cite{PhysRevA.79.063605}.  A mixture of atoms in the $\left|F=9/2, m_F=9/2\right\rangle$ and $\left|F=9/2, m_F=7/2\right\rangle$ hyperfine states is used for initial forced evaporative cooling in a variation of the QUIC magnetic trap \cite{Esslinger1998}.  Transitions from both spin states to magnetically un-trapped Zeeman levels in the  $F=7/2$ hyperfine state are driven using a single microwave-frequency magnetic field. After cooling in the QUIC trap, the atoms are transferred into a crossed-beam 1064~nm dipole trap.  Following additional cooling induced by reducing the laser intensity, atoms in the $\left|F=9/2, m_F=7/2\right\rangle$ state are removed using microwave transitions and a magnetic field gradient. The number of atoms in the gas ranges from $6\times10^4$ to $3\times10^5$ for all of the data in this manuscript, and the measured spin polarization is greater than 65:1. The waist of the Gaussian beam used for the dipole trap is approximately 120~$\mu$m, and the trap frequencies vary from $\omega_z=\omega_y=2\pi\times$120~Hz, $\omega_x=2\pi\times50$~Hz to $\omega_z=\omega_y=2\pi\times$210~Hz, $\omega_x=2\pi\times80$~Hz during the final stage of cooling.  The final temperature of the gas is correlated with the trap frequencies, with the lowest (highest) temperature corresponding to the lowest (highest) trap frequencies.

The optical speckle field is generated by passing a 532~nm laser beam with a 16~mm waist through a 15~mm diameter plano-convex lens with a 13~mm focal length (Lightpath Industries GPX-15-15).  A 0.25~mm thick holographic diffuser (from Luminit, LLC) mounted to the flat side of the focusing lens scatters the light through a 0.5~degree range of angles.  The waist of the speckle field envelope is determined in situ by measuring the dipole force from the envelope on the mobile component, as in \cite{White2009}.  We determine $\Delta$ within a 10\% systematic uncertainty by calculating the average intensity based on the measured waist and 532~nm laser power.  The Rayleigh length that characterizes the envelope along $z$ is measured ex-situ.

Temperatures are determined by measuring the expansion velocity of the gas without disorder present.  The temperature and corresponding uncertainty are calculated from a linear fit to the r.m.s. size measured at different free expansion times.  We determine the maximum impact of disorder on the temperature and momentum distribution using approximately the lowest temperature and highest disorder energy that we sample.  We prepare a gas at $\left(262\pm4\right)$~nK in the dipole trap, then turn on the speckle potential, and then measure the momentum distribution after simultaneously turning off the speckle potential and the trap. For $\Delta\approx800$~$k_B\times nK$, we determine that the temperature of the gas is $\left(288\pm9\right)$~nK.  Thus, the speckle potential induces at most a 10\% shift in the temperature.  We also verify that the disorder does not affect the shape of the momentum distribution using a chi-squared analysis of the Gaussian fit to the time-of-flight images.  We find that the reduced chi-squared is $1.26\pm0.22$ with $\Delta=0$~$k_B\times nK$ and $1.32\pm0.25$ for $\Delta\approx800$~$k_B\times nK$.  The disorder therefore leads to no significant change in the shape of the momentum distribution.

Images of the gas after expansion in the speckle potential are analyzed by first fitting only the mobile component to a Gaussian profile. This is accomplished by excluding the region containing the localized component using a rectangular mask. The size of the mask is approximately twice the size of the localized component along $x$ and runs across the entire image in $z$.  Small changes in the width of the mask do not affect the measured localized fraction, $\xi_z$, or $\sigma_x$. The mobile component is removed from the image by subtracting the Gaussian fit, and the residual localized component is fit to a function proportional to $e^{-x^2/2\sigma_x^2-|z|/\xi_z}$. The centers of the fits to the mobile and localized components are treated as independent free parameters. The localized fraction is determined from the fitted integrated column density in each component.  Fringes that arise from technical noise are removed from the images shown in Figs. 1--3 using standard image processing techniques; this procedure is not applied to images used for determining the properties of the localized component.

The simulation of classical trajectories is carried out for a range of particle energies, averaging over randomly sampled initial position. The computed dynamics are diffusive along all directions for expansion times greater than a few ms.  To simulate the expansion of the gas, we average the diffusion constants over a thermal ensemble of energies consistent with the initial momentum distribution.  The simulated r.m.s. radius along $z$ is converted to a localization length for comparison to the data by assuming an exponential profile.  The motion in $x$ and $y$ reaches the asymptotic regime within 10~$\mu$m in $z$.  Therefore, for the localization lengths we measure, localization solely in $z$ cannot explain the absence of diffusion in the transverse directions given the simulated diffusion rates.

We determined that the speckle envelope does not directly affect the measured scaling of the mobility edge by analyzing the raw data from Fig. 3 using a different method.  The disorder energy $\overline{\Delta}$ averaged across the density profile is computed using the measured dimensions of the gas.  For this alternative analysis, $\overline{\Delta}$ is employed as a measure of the disorder strength instead of $\Delta$. The mobility edge is computed for each point (at a specific temperature and speckle intensity) and plotted vs. $\overline{\Delta}$.  The data are fitted to a power law, which gives $E_c\propto\overline{\Delta}\;^{0.58\pm0.03}$, in agreement with the simpler analysis shown in Fig. 3B.

\clearpage

\bibliography{toscience}

\begin{thebibliography}{10}
\bibitem{SOM}
$\textrm{Materials and methods are available as supporting material}$.
\end{thebibliography}

\bibliographystyle{Science}

\begin{scilastnote}
\item[] We thank L. Sanchez-Palencia for stimulating discussions and M. White and P. Koehring for technical assistance.  We acknowledge funding from the DARPA OLE program, Office of Naval Research (award \#N000140911023), and the National Science Foundation (award \#0855027).
\end{scilastnote}

\end{document}